\documentclass[12pt,epsfig]{article}
\usepackage[dvips]{graphicx}

\def\lromn#1{\uppercase\expandafter{\romannumeral#1}}

\begin{document}

\begin{titlepage}

\begin{center}

\hfill KEK-TH-1046\\
\hfill TUM-HEP-607/05\\
\hfill SISSA-81/2005/EP\\
\hfill \today

\vspace{1cm}
{\large\bf Effective theoretical approach of Gauge-Higgs unification
model and its phenomenological applications}
\vspace{1cm}

{\bf Naoyuki Haba}$^{(a,b),}$
\footnote{E-mail: haba@ph.tum.de},
{\bf Shigeki Matsumoto}$^{(c),}$
\footnote{E-mail: smatsu@post.kek.jp},
{\bf Nobuchika Okada}$^{(c,d),}$
\footnote{E-mail: okadan@post.kek.jp},
and
{\bf Toshifumi Yamashita}$^{(c,e),}$
\footnote{E-mail: tyamashi@post.kek.jp}
\vskip 0.15in
{\it
$^{(a)}${Institute of Theoretical Physics, University of
Tokushima, 770-8502, Japan}\\
$^{(b)}${Physik-Department, Technische Universitat Munchen,
James-Franck-Strasse, D-85748 Garching, Germany}\\
$^{(c)}${Theory Group, KEK, Oho 1-1 Tsukuba, 305-0801, Japan}\\
$^{(d)}${The Graduate University for Advanced Studies (Sokendai), Oho 1-1 Tsukuba, 305-0801, Japan}\\
$^{(e)}${SISSA, Via Beirut 2, I-34014 Trieste, Italy}
}
\vskip 1in

\abstract{We derive the low energy effective theory of Gauge-Higgs
unification (GHU) models in the usual four dimensional framework. We find
that the theories are described by only the zero-modes with a particular
renormalization condition in which essential informations about GHU models
are included. We call this condition ``Gauge-Higgs condition'' in this letter.
In other wards, we can describe the low energy theory as the SM with this
condition if GHU is a model as the UV completion of the Standard Model. This
approach will be a powerful tool to construct realistic models for GHU and to
investigate their low energy phenomena.}

\end{center}
\end{titlepage}
\setcounter{footnote}{0}

\vspace{1.0cm}
\lromn 1 \hspace{0.2cm} {\bf Introduction}
\vspace{0.5cm}

The hierarchy problem in the Standard Model (SM) is expected to give the clue
to explore the physics beyond the SM. The problem is essentially related to
the quadratically divergent corrections to the Higgs mass, which reduce the
predictive power of the model. To avoid the divergence, many scenarios have
been proposed so far: for example, supersymmetry, TeV scale extra dimension 
\cite{TeVXD},
and so on. Recently the models based on the Gauge-Higgs unification (GHU)
scenario \cite{Manton:1979kb,YH} attract attentions for solving the problem
\cite{YH}-\cite{Hosotani:2004ka}. In the scenario, the models are defined in
the higher dimensional space-time in which the extra dimensions are
compactified on an appropriate orbifold. The Higgs field is then identified
as the zero mode of the extra dimensional components of the gauge field.
Since the gauge invariance in the higher dimension protects the Higgs
potential from ultraviolet (UV) divergences, we can avoid the hierarchy
problem.

One important prediction of the models is the Higgs potential (or Higgs mass
and its interactions as physical observables), because almost all
interactions are governed by the gauge invariance. While the potential
vanishes at tree level due to the invariance, it is produced from radiative
corrections induced from the compactification. Therefore, we have to
calculate at least 1-loop effective potential for the Higgs field. Though it
has been achieved in some simple models such as the toroidal compactification
\cite{Lim}, it is an awkward task in more generic case and/or at higher-loop
levels.

This task will be much easier if we can construct the low energy effective
theory in the four dimensional view point. In this letter, we show that the
construction is possible for GHU models. This fact is supported from the
discussion in Ref.\cite{Takenaga:2003dp}, in which it is argued that the
effective potential is governed by the infrared (IR) physics.

The effective theories are described by the zero modes in the usual four
dimensional framework. Since the Higgs field is merely a scalar field in the
effective theories, its potential receives the divergent corrections. Thus we
have to renormalize the potential. The main result in this work is that the
Higgs potential calculated in the original model is reproduced when we
require the particular renormalization condition. The condition is settled at
the scale $1/2\pi R$, where $R$ is the radius of the extra dimension. To be
more precise, we impose the running coupling constant for the self coupling
of the Higgs filed $\lambda$ becomes zero at that scale,
$\lambda(1/2\pi R) = 0$. We call this condition {\it Gauge-Higgs condition}.
The condition can be intuitively understood, because the effective theory is
matched with the GHU model itself at the compactification scale, in which the
interaction must be vanished. It means we can describe the low energy theory
as the SM (+ possible vector-like fermions) with the Gauge-Higgs condition if
a GHU model is realized as the UV completion of the SM. The situation for
imposing the condition at the cutoff scale is similar to that in the top
condensation model, where the model is effectively identified as the SM with 
the so-called compositeness condition imposed at a cutoff
(composite) scale \cite{BHL}.
 
Since we are very familiar with the treatment of the usual four dimensional
field theory, the low energy effective theory will be a powerful tool to
construct realistic models in the GHU scenario and to investigate their low
energy phenomena. For example, we can use the effective theory to construct
the GHU models that reproduce the Standard Model correctly. Or we can obtain
a renormalization group (RG) improved analysis for the Higgs mass, which is
difficult to make in the original higher dimensional framework due to the
unrenormalizability of the theory. 

This letter is organized as follows. In the next section, we briefly
introduce the GHU scenario using a simple toy model defined in five
dimensions. We calculate the effective potential of the Higgs field in terms
of the five dimensional view point in the end of this section. In section
\lromn 3, we construct a low energy effective theory and show that the
potential derived in the previous section is reproduced with the use of the
Gauge-Higgs condition. After the construction, we discuss some applications
to low energy phenomena using the effective theory in section \lromn 4.
Section \lromn 5 is devoted to summary.

\vspace{1.0cm}
\lromn 2 \hspace{0.2cm} {\bf GHU scenario and Higgs potential}
\vspace{0.5cm}

We briefly review the GHU scenario using a toy model in this section. We
especially focus on how the doublet Higgs field is produced from the higher
dimensional gauge field, and on the mass spectrum of the Kaluza-Klein (KK)
particles. Finally, we discuss the Higgs potential at 1-loop level in the
toy model, which will be compared with the potential calculated in the 
framework of the low energy effective theory.

\newpage

\vspace{0.5cm}
\underline{\bf Five dimensional SU(3) model}
\vspace{0.5cm}

We use the five dimensional SU(3) model for explaining the GHU scenario and
for discussing the Higgs potential. Though the model is regarded as a toy
model because it does not yields the correct Weinberg angle, it is sufficient
to use the model for our purpose. Application to more realistic models is
straightforward.

The model is given by the Yang-Mills model defined on the five dimensional
space-time in which the fifth direction, $y$, is compactified on the orbifold
$S^1/Z_2$. The particle contents are the five dimensional gauge field
$A_a(x^\mu,y)$ and bulk fermions $\Psi(x^\mu,y)$, where the subscript $a$
runs from 0 to 3 and 5. Due to the compactification, the action must be
invariant under two operations, those are the translation T : $y$ to
$y + 2\pi R$ and the parity P : $y$ to $-y$. The radius of the circle $S^1$
is denoted by $R$. 

At first, we discuss the gauge boson sector of this model. Under the
operations T and P, the four dimensional component of the gauge field
$A_\mu$ and the fifth one $A_5$ are set to be transform as 
\begin{eqnarray}
 &&
 A_\mu(x^\mu, y + 2\pi R)
 =
 \hat{T}^\dagger A_\mu(x^\mu, y)\hat{T}~,
 \qquad
 A_\mu(x^\mu, - y)
 =
 \hat{P}^\dagger A_\mu(x^\mu, y)\hat{P}~,
 \nonumber \\
 &&
 A_5(x^\mu, y + 2\pi R)
 =
 \hat{T}^\dagger A_5(x^\mu, y)\hat{T}~,
 \qquad
 A_5(x^\mu, -y)
 =
 -\hat{P}^\dagger A_5(x^\mu, y)\hat{P}~,
\end{eqnarray}
where the operator $\hat{T}$ and $\hat{P}$ are defined by
$\hat{T}$ = diag(1, 1, 1) and $\hat{P}$ = diag($-1$, $-1$, 1). By these
boundary conditions, the SU(3) gauge symmetry is broken into
SU(2)$\times$U(1) symmetry \cite{Kawamura}. In terms of SU(2)$\times$U(1),
the gauge fields are decomposed as 
\begin{eqnarray}
 &&
 A_\mu~~:~~
 {\bf 8}
 ~~\rightarrow~~
 {\bf 3}_0^{(+,+)}
 +
 {\bf 2}_{1/2}^{(+,-)}
 +
 {\bf 2}_{-1/2}^{(+,-)}
 +
 {\bf 1}_0^{(+,+)}~,
 \nonumber \\
 &&
 A_5~~:~~
 {\bf 8}
 ~~\rightarrow~~
 {\bf 3}_0^{(+,-)}
 +
 {\bf 2}_{1/2}^{(+,+)}
 +
 {\bf 2}_{-1/2}^{(+,+)}
 +
 {\bf 1}_0^{(+,-)}~,
 \label{adjoint}
\end{eqnarray}
where the lower subscripts represent the hyper charges of U(1) gauge
interaction, and the upper ones are charges for T and P operations. Since
only components with $(+,+)$ have zero-modes, we can confirm that the SU(3)
symmetry is in fact broken to SU(2)$\times$U(1). Furthermore the zero-modes
of $A_5$ behaves as a doublet scalar so that we can identify these particles
as the SM Higgs doublet.

According to the method proposed in Ref.\cite{GeneralFormula}, we can 
calculate the mass eigenvalues of the KK particles $m_{\rm Gauge}$, which
are used to calculate the Higgs potential. After some calculations, we obtain
\begin{eqnarray}
 \left\{
  m_{\rm Gauge}^2
 \right\}
 =
 \left\{
  \left(n + Q_i^{(G)} a\right)^2/R^2
 \right|
 \left.
  n \in Z~,
  ~~~
  Q_i^{(G)}
  =
  0,~-\frac{1}{2},~\frac{1}{2},~1
  \right\}~.
\end{eqnarray}
In the above formula, we take the effect of the spontaneous symmetry breaking
of the Higgs field into account. The vacuum expectation value (VEV) of $A_5$
is denoted by $a$. To be more precise, it is defined by
$\langle A_5\rangle = a\lambda^4/(2gR)$, where $\lambda^4$ is the 4th
Gell-Mann matrix and $g$ is the bulk gauge coupling. The relation between $g$
and four dimensional gauge coupling $g_4$ is given by $g_4 = \sqrt{2\pi R}g$
and the weak scale VEV $v \sim$ 246 GeV is written as $a = g_4Rv$.

Next we discuss the fermion sector of the model. We consider the case that
the bulk fermion is belonging to the fundamental $\Psi_F$ or adjoint
representation $\Psi_A$. Under the operation of T and P, these fermions
transform as 
\begin{eqnarray}
 &&
 \Psi_F(x^\mu, y + 2\pi R)
 =
 \xi \hat{T}\Psi(x^\mu, y)~,
 \qquad
 \Psi(x^\mu, -y)
 =
 \eta\gamma_5\hat{P}\Psi_F(x^\mu, y)~,
 \nonumber \\
 &&
 \Psi_A(x^\mu, y + 2\pi R)
 =
 \xi\hat{T}^\dagger \Psi_A(x^\mu, y)\hat{T}~,
 ~~~
 \Psi(x^\mu, -y)
 =
 \eta\gamma_5\hat{P}^\dagger \Psi_A(x^\mu, y)\hat{P}~,
\end{eqnarray}
where $\xi$, $\eta$ denote the overall signs which can be $+$ or $-$, and
$\gamma_5$ is the chirality operator. Thus we have four kinds of fermions in
both fundamental and adjoint representations due to the choice of $\eta$ and
$\xi$. These representations of SU(3) are decomposed in terms of
SU(2)$\times$U(1) as 
\begin{eqnarray}
 &&
 {\bf 3}(\xi,\eta)
 ~~\rightarrow~~
 {\bf 2}_{1/6}^{(\xi,-\eta)}
 +
 {\bf 1}_{-1/3}^{(\xi,\eta)}~,
 \nonumber \\
 &&
 {\bf 8}(\xi,\eta)
 ~~\rightarrow~~
 {\bf 3}_0^{(\xi,\eta)}
 +
 {\bf 2}_{1/2}^{(\xi,-\eta)}
 +
 {\bf 2}_{-1/2}^{(\xi,-\eta)}
 +
 {\bf 1}_0^{(\xi,\eta)}~.
\end{eqnarray}
The dependence of $\xi$ and $\eta$ mean that once the $\eta$ and $\xi$ are
fixed in the left side of the formula, the charges for T and P operations
described by the upper subscript in the right side are determined. The bulk
fermions with a positive T charge (periodic condition) have zero-modes, while
those with a negative T (anti-periodic condition) does not have.

As in the case of the gauge fields, we can calculate the mass eigenvalues
of KK fermions. The eigenvalues for T-even and T-odd fermions in the cases
of fundamental and adjoint representations $m_{\rm Fund^+.}$,
$m_{\rm Fund^-.}$, $m_{\rm Adjo^+.}$ and $m_{\rm Adjo^-.}$ are written as
\begin{eqnarray}
 \left\{
  m_{\rm Fund^+.}^2
 \right\}
 &=&
 \left\{
  \left(n + Q_i^{(F^+)} a\right)^2/R^2 + m_{F^+}^2
 \right|
 \left.
  n \in Z,
  ~
  Q_i^{(F^+)}
  =
  0, -\frac{1}{2}, \frac{1}{2}
 \right\}~,
 \\
 \left\{
  m_{Adjo^+.}^2
 \right\}
 &=&
 \left\{
  \left(n + Q_i^{(A^+)} a\right)^2/R^2 + m_{A^+}^2
 \right|
 \left.
  n \in Z,
  ~
  Q_i^{(A^+)}
  =
  0, -\frac{1}{2}, \frac{1}{2}, 1
 \right\}~,
 \nonumber \\
 \left\{
  m_{\rm Fund^-.}^2
 \right\}
 &=&
 \left\{
  \left(n + Q_i^{(F^-)} a + \frac{1}{2}\right)^2/R^2 + m_{F^-}^2
 \right|
 \left.
  n \in Z,
  ~
  Q_i^{(F^-)}
  =
  0, -\frac{1}{2}, \frac{1}{2}
 \right\}~,
 \nonumber \\
 \left\{
  m_{Adjo^-.}^2
 \right\}
 &=&
 \left\{
  \left(n + Q_i^{(A^-)}a + \frac{1}{2}\right)^2/R^2 + m_{A^-}^2
 \right|
 \left.
  n \in Z,
  ~
  Q_i^{(A^-)}
  =
  0, -\frac{1}{2}, \frac{1}{2}, 1
 \right\}~,
 \nonumber
\end{eqnarray}
where the mass terms $m_{F^+}$, $m_{F^-}$, $m_{A^+}$ and $m_{A^-}$ arising in
the right side of above formulas represent the bulk mass terms in each
fermion. For the adjoint representation, the spin degree of freedom is four,
while it is two for the fundamental representation.

\vspace{0.5cm}
\underline{\bf Effective potential}
\vspace{0.5cm}

In this subsection, we discuss the effective potential of the Higgs field in
the five dimensional SU(3) model. The calculation at 1-loop level is
completely performed in Ref.\cite{GeneralFormula}, thus we show only the
result here. The important result in the calculation is that the potential
does not suffer from a UV divergence due to the higher dimensional gauge
invariance and we obtain the finite result without renormalizations. In other
words, the finiteness of the potential comes from the existence of KK
particles. Though the calculation of the potential by using only zero-modes
leads to a UV divergence, it disappears after summing up all KK modes. It means
that the the physical cutoff at loop integrations is naturally provided by
the summation. After some calculation, the potential turns out to be
\begin{eqnarray}
 V(\phi)
 &=&
 -L\frac{C}{2}
 \sum_i\sum_{w = 1}^\infty
 \frac{1}{w^5}
 \left[
  3\cos(2\pi wQ_i^{(G)}\phi)
  \rule{0mm}{0.7cm}
 \right.
 \label{Exact 1}
 \\
 &-&
 2\cos\left(2\pi wQ_i^{(F^+)}\phi\right)\sum_{j = 1}^{N_{F^+}}
 \left(
  1 + wLm_{F^+}^j + \frac{(wLm_{F^+}^j)^2}{3}
 \right)
 e^{-wLm_{F^+}^j}
 \nonumber \\
 &-&
 4\cos\left(2\pi wQ_i^{(A^+)}\phi\right)\sum_{j = 1}^{N_{A^+}}
 \left(
  1 + wLm_{A^+}^j + \frac{(wLm_{A^+}^j)^2}{3}
 \right)
 e^{-wLm_{A^+}^j}
 \nonumber \\
 &-&
 2\cos\left(2\pi w\left(Q_i^{(F^-)}\phi - \frac{1}{2}\right)\right)
 \sum_{j = 1}^{N_{F^-}}
 \left(
  1 + wLm_{F^-}^j + \frac{(wLm_{F^-}^j)^2}{3}
 \right)
 e^{-wLm_{F^-}^j}
 \nonumber \\
 &-&
 \left.
  4\cos\left(2\pi w\left(Q_i^{(A^-)}\phi - \frac{1}{2}\right)\right)
  \sum_{j = 1}^{N_{A^-}}
  \left(
   1 + wLm_{A^-}^j + \frac{(wLm_{A^-}^j)^2}{3}
  \right)
  e^{-wLm_{A^-}^j}
 \right]~,
 \nonumber
\end{eqnarray}
up to the constant term. The field $\phi$ is the rescaled $A_5$ field defined
by $A_5 = \phi\lambda^4/(2gR)$, thus its vacuum expectation value is given
by $a$. The relation between the field $\phi$ and the Higgs field $h$ is given
by $\phi = g_4Rh$. The coefficient $C$ is $C = 3/(2\pi^2(2\pi R)^5)$ and
$L = 1/2\pi R$ is the length of circumference of the extra dimension. The
factor $L$ in the front comes from the integration of the 5th direction. The
parameters $N_{F^\pm}$ and $N_{A^\pm}$ are the number of fundamental and
adjoint fermions included in the model. Fundamental fermions must be
introduced with a pair due to the gauge anomaly cancellation, thus $N_{F^\pm}$
is even number. The physical interpretation of $w$ in the equation is the
winding number of the internal loop along with the $S^1$ direction. The terms
in the parenthesis correspond to the contributions from gauge bosons,
fundamental and adjoint periodic fermions, and those of anti-periodic
fermions, respectively.

The shift $-1/2$ of the cosine function in the contributions from
anti-periodic fermions yields an additional sign factor $(-1)^w$ compared to
contributions form periodic fermions. The factor is induced from the
anti-periodicity of the loop integrals along with the extra dimension. Even
though we introduce only bulk fermions, we can obtain both positive and
negative mass squared corrections in the potential. Therefore we can
construct models where the quadratic term of the Higgs field is negative and
very small compared to the compactification scale due to the cancellation
between these contributions. As a result, we obtain a small VEV of the Higgs
field without introducing any scalar fields.

The weak scale VEV is expected to be small enough compared to the
compactification scale $1/R$, thus we expand the potential by the field
$\phi$ and express it as a power series of the field. This form is used to
compare the result from the potential obtained from the low energy effective
theory in the next section. We also assume that the bulk masses of fermions
are small, otherwise the contributions to the effective potential from
these particles are negligible due to the exponential factor in
Eq.(\ref{Exact 1}), namely they are decoupled from the effective theory. 
So we focus on first few terms of $\phi$ and $m$ in the expansion. After
expansion, the potential in Eq.(\ref{Exact 1}) is written as
\begin{eqnarray}
 V(\phi)
 &=&
 \frac{F_2}{2}(2\pi\phi)^2
 +
 \frac{F_4(\phi)}{4!}(2\pi\phi)^4
 +
 {\cal O}(\phi^6)~,
\end{eqnarray}
where the coefficients $F_2$ and $F_4$ are defined as
\begin{eqnarray}
 F_2
 &=&
 L\frac{C}{2}\zeta_R(3)\sum_i
 \left[
  \sum_{r = G,F^+,A^+}
  d_rN_r(Q_i^{(r)})^2
  -
  \sum_{r = F^-,A^-}
  \frac{3}{4}d_rN_r(Q_i^{(r)})^2
 \right]~,
 \label{phi2}
 \\
 \nonumber \\
 F_4(\phi)
 &=&
 L\frac{C}{4}\sum_i
 \left[
  ~~\sum_{r = G,F^+,A^+}
  d_rN_r(Q_i^{(r)})^4
  \left(
   \ln\left\{(2\pi Q_i^{(r)}\phi)^2\right\}
   -
   \frac{25}{6}
  \right)
 \right.
 \nonumber \\
 &&
 \qquad\qquad
 \left.
  +\sum_{r = F^-,A^-}
  d_rN_r(Q_i^{(r)})^4\ln 4
 \right]~,
 \label{phi4}
\end{eqnarray}
where $\zeta_R(x)$ is Riemann's zeta function. In the expansion, we omit the
constant terms, namely the contributions to the cosmological constant. In 
Eqs.(\ref{phi2}) and (\ref{phi4}), the parameters $d_r$ are the spin degree
of freedom and defined as $d_G = 3$, $d_{F^\pm} = -2$ and $d_{A^\pm} = -4$.
The number of the gauge boson is of course one ($N_G = 1$). In the expansion,
we neglect the bulk mass terms of fermions for simplicity. For the detailed
formulas including the mass terms, refer to Appendix.

\vspace{1.0cm}
\lromn 3 \hspace{0.2cm} {\bf Low energy effective theory of GHU models}
\vspace{0.5cm}

We construct the low energy effective theory of GHU models describing the
physics at the scale lower than $1/R$ in this section. For this purpose, we
use the five dimensional SU(3) model again. The effective theory must be
described by only zero modes in the four dimensional space-time because the
masses of higher modes are of order $1/R$ and they are already integrated
out. In the SU(3) model, the particle contents in the effective theory are
SU(2) and U(1) gauge fields, Higgs field and zero modes of bulk fermions. The
interactions between these particles are uniquely determined by the original
five dimensional SU(3) model. As shown in the previous section, the self 
interactions of the Higgs field are induced from the radiative corrections
through the compactification, thus it is not trivial to write them down.
Therefore we derive the interactions from the comparison between the Higgs
potential calculated in the effective theory and that from the original SU(3)
model. 

The Higgs potential obtained from the calculation in the effective theory has
UV divergences, it should be renormalized with an appropriate renormalization
condition. By the comparison mentioned above, we can fix the condition and
derive the Higgs interactions. Namely we can obtain the matching condition
between the effective theory and the original SU(3) model by the comparison.

We consider the mass term of the Higgs field. As shown in Eq.(\ref{phi2}),
there are contributions from periodic modes ($F^+$, $A^+$) and anti-periodic
modes ($F^-$, $A^-$) in addition to the term from gauge bosons. Both
corrections are regularized by the compactification scale thanks to the
higher dimensional gauge invariance. An important point is that the mass
corrections from anti-periodic modes are of the same order as that from a
periodic mode but have opposite sign. Furthermore the anti-periodic fermions
have no zero modes and its contribution to the Higgs self coupling is
suppressed compared to that from the periodic ones. Thus we can tune the mass
parameter by introducing anti-periodic fermions without altering the low
energy effective theory. This fact means that we can treat the mass parameter
as a free parameter as far as we are interested in only the low energy
phenomenology of GHU models.

Next we discuss the self coupling of the Higgs field. There are several
contributions to the coupling. Among those, the contributions from the
anti-periodic fermions are small compared to other contribution as can be
seen in Eq.(\ref{phi4}). Thus we neglect these terms in our discussion and
focus on the contributions from gauge bosons and periodic fermions. At first,
we consider the contribution from the periodic and fundamental fermion
without the bulk mass for simplicity. From Eq.(\ref{phi4}), the contribution
is rewritten in terms of Higgs field ($\phi = g_4Rh$),
\begin{eqnarray}
 \left.V^{(F^+)}(h)\rule{0mm}{0.4cm}\right|_{h^4}
 &=&
 -\frac{1}{4!}\frac{3g_4^4}{32\pi^2}
 \left[
  \ln\left(\frac{(g_4hL)^2}{4}\right)
  -
  \frac{25}{6}
 \right]h^4
 \nonumber \\
 &\simeq&
 -\frac{1}{4!}\frac{3g_4^4}{32\pi^2}
 \left[
  \ln\left(\frac{h^2}{1/L^2}\right)
  -
  \frac{25}{6}
 \right]h^4~,
 \label{eff1}
\end{eqnarray}
where $h$ is the real neutral component of the doublet scalar $H$, that is
the zero-mode of $A_5$ and defined by $\phi = g_4Rh$. At the last equation,
we have neglected  the term $\ln (g_4^2/4)$, because it is small enough
compared to other terms for $g_4\sim 1$. For the effect of this term, refer
to the discussion in the end of this section.

The corresponding contribution to the Higgs field is calculated in the
framework of the low energy effective theory. The zero-modes of these
fermions multiplet consist of a doublet $\Psi_L$ and a singlet $\Psi_R$ as
zero-modes. They have the following Yukawa coupling
with $H$: 
\begin{eqnarray}
 {\cal L}_{\rm Yukawa}
 =
 \frac{g_4}{\sqrt{2}}
 \bar{\Psi}_L\Psi_R H + h.c.~.
\end{eqnarray}
Using the Yukawa interactions, we can calculate the contributions to the
Higgs potential $V^{(F^+)}_{\rm eff.}(h)$. As mentioned above, the correction
has UV divergences which should be renormalized with an appropriate
renormalization condition. Since we can not define a renormalized self
coupling around the origin ($h=0$) due to the IR divergences, we define the
coupling at a non-vanishing renormalization point $\mu$ as adopted in the
reference \cite{Coleman},
\begin{eqnarray}
 \left.\frac{{\rm d}^4V_{\rm eff}}{{\rm d}h^4}\right|_{h = \mu}
 =
 \lambda(\mu)~.
\end{eqnarray}
With this renormalization condition, the contribution to the Higgs potential
from the fundamental fermions is written as
\begin{eqnarray}
 \left.V_{\rm eff}^{(F^+)}\rule{0mm}{0.4cm}\right|_{h^4}
 =
 \frac{1}{4!}
 \left[
  \lambda(\mu)
  +
  \frac{b}{2}\left(\frac{g_4}{2}\right)^4
  \left\{
   \ln\left(\frac{h^2}{\mu^2}\right) - \frac{25}{6}
  \right\}
 \right]h^4~,
 \label{eff2}
\end{eqnarray}
where $b= -3/\pi^2$ is the coefficient of the beta function concerning the
Yukawa coupling. The running coupling $\lambda(\mu)$ obeys the following
renormalization group (RG) equation,
\begin{eqnarray}
 \frac{d\lambda}{d\ln\mu}
 =
 b\left(\frac{g_4}2\right)^4~.
 \label{RG massless}
\end{eqnarray}
By comparing Eq.(\ref{eff1}) with Eq.(\ref{eff2}), we find 
the renormalization condition,
\begin{eqnarray}
 \lambda\left(\frac{1}{L}\right)
 =
 \lambda\left(\frac{1}{2\pi R}\right)
 =
 0~.
 \label{G-H condition}
\end{eqnarray}
This is the ``Gauge-Higgs condition'' mentioned in Introduction. The other
contributions to the Higgs potential from gauge bosons and adjoint fermions
are calculated in the same manner, and we achieve the same result that the
contributions obtained in the original GHU model can be reproduced by
imposing the Gauge-Higgs condition.

We comments on the effects of bulk masses of fermions. Again we use the
fundamental fermion with the periodic condition as an example. From
Eq.(\ref{Exact 1}), the contribution to the Higgs potential with the bulk
mass is written as
\begin{eqnarray}
 \left.V^{(F^+)}(h)\rule{0mm}{0.4cm}\right|_{h^4}
 &=&
 -\frac{1}{4!}\frac{3g_4^4}{32\pi^2}
 \left[
  \ln\left(\frac{(g_4hL)^2}{4} + L^2m_{F^+}^2\right)
  -
  \frac{25}{6}
 \right]h^4~,
 \label{docoupling}
\end{eqnarray}
where we assume that the vacuum expectation value $h$ and the bulk mass
$m_{F^+}$ are small enough compared to the compactification scale $1/R$, and
use the expansion formula discussed in Appendix. When we consider the running
coupling $\lambda(\mu) = d^4V(\mu)/d\mu^4$ in this case, it should be
coincide with the one without the bulk mass in the range $\mu \gg m$. Since
we now consider the situation $m_{F^+} \ll 1/R$, we obtain the Gauge-Higgs
condition to reproduce the potential again. The difference appears at the
scale smaller than the mass $\mu \ll m$. As can be seen in the above formula,
the coupling does not move due to the mass term in this range. This is the
decoupling phenomenon, thus the effect can be taken into account by using the
equation,
\begin{eqnarray}
 \frac{d\lambda_m}{d\ln\mu}
 =
 b\left(\frac{g_4}2\right)^4\theta(\mu - m_{F^+})~.
\end{eqnarray}
with the Gauge-Higgs condition $\lambda(1/L) = 0$ in stead of that in
Eq.(\ref{RG massless}).

Here we discuss the term $\ln(Q_i^2g_4^2)$, which has been neglected in
Eq.(\ref{eff1}). After the symmetry breaking (the Higgs field gets the
vacuum expectation value $v$), the argument of the logarithm in
Eq.(\ref{docoupling}) is written as
$(Q_i^2g_4^2v^2 + m_{F^+}^2)/(1/L^2)$, $Q_i = 1/2$. The numerator of this 
formula is nothing but the physical mass of the bulk fermion. Thus the
argument of the logarithm represents the running effect of the quartic
coupling $\lambda$ between the UV cut off scale ($1/L$) and the IR cut off
scale (physical mass). In this meaning, the term $\ln(Q_i^2g_4^2)$ has an
important role for describing the decoupling phenomenon, though its effect
is practically negligible compared to other terms. If we consider the
effective potential in the low energy effective theory in more detail,
for example, considering threshold corrections, the Higgs potential for GHU
models may be reproduced more precisely and include the $\ln(Q_i^2g_4^2)$.
We leave this problem as a future work.

The Gauge-Higgs condition is that all GHU models should satisfy. Thus, if we
construct a realistic GHU model, its effective theory should be the SM
(+ possible association with massive vector-like fermions) with this
condition as the boundary condition of RG flow. Once we clarify the feature
of the effective theory in the GHU scenario, we are able to analyze the GHU
models by using the effective theory. This reduces the necessary efforts
greatly. We will show some examples of applications in the next section. 

\vspace{1.0cm}
\lromn 4 \hspace{0.2cm} {\bf Application to phenomenology}
\vspace{0.5cm}

The low energy effective theory we have developed will be a powerful tool to
construct the realistic model in the GHU scenario and to investigate their
low energy phenomena. In this section, we apply the effective theory to a
phenomenology.

As mentioned in the previous section, the Gauge-Higgs condition will be
imposed in all GHU models. In the construction of the realistic model, 
we have some troubles in general. For instance, it is difficult to reproduce
a realistic top Yukawa coupling in GHU models where all Yukawa couplings is
written by the SU(2) gauge coupling at the compactification scale. It would
require a somewhat complicated set-up to yields a large top Yukawa coupling 
\cite{5DLorentzV,LargeRepr}. 
Even if we can construct the realistic model, it may be a hard task to
calculate some low energy observables such as a Higgs potential in an
original extra dimensional model.

On the other hand, from the viewpoint of the low energy physics, the
effective theory of the realistic models should be described by the SM with
the Gauge-Higgs condition. 
In fact, if we introduce the setups proposed in 
 Refs\cite{5DLorentzV,LargeRepr} to explain the large top Yukawa coupling, 
 we can show the condition holds. 
In addition, even if we consider more complicated extensions, for example 
 models where an additional U(1) gauge symmetry is imposed to reproduce 
 a correct Weinberg angle, the fact that 
the Higgs potential vanishes at the compactification scale will be 
unchanged, as far as the Higgs field corresponds to the degree of freedom of 
the Wilson line. 
This is because above the scale, the Higgs fields behaves as the Wilson line 
which has vanishing potential.
Note that there is a possibility that the scale to impose
the condition is modified slightly due to the details of setups.
The scale is , however, much higher than the weak scale, and its correction 
has little effects on the Higgs mass.
It means that we can investigate the low energy
phenomena without detailed informations about the models. By using the
advantage, we make an RG improved analysis of the Higgs mass in the GHU
scenario in the following.

\begin{figure}[t]
 \begin{center}
  \scalebox{0.8}{\includegraphics*{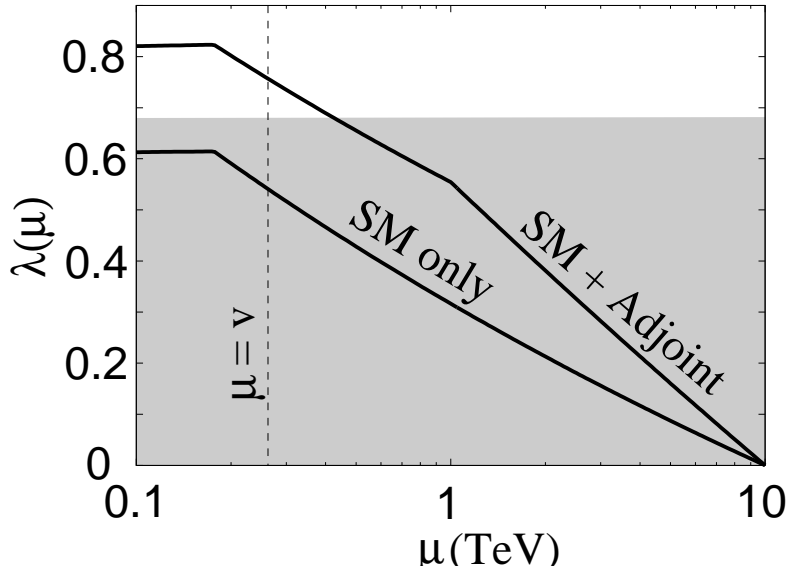}}
  \caption{\small RG flows of the Higgs quartic coupling at 1-loop level. The
  UV cutoff scale $1/L$ is set to be 10 TeV. The shaded area corresponds to
  the LEP bound ($m_h \ge$ 114 GeV). The vertical line shows
  $\mu = \langle h\rangle =$ 246 GeV where the Higgs mass should be evaluated.
  The lower flow is for the case of the SM contribution only. The upper flow
  includes an additional fermion pairs in the adjoint representation of SU(3)
  with bulk mass of 1 TeV.}
  \label{fig:lambda}
 \end{center}
\end{figure}

The RG equations for SM interactions at 1-loop level \cite{RGE} are given by
\begin{eqnarray}
 \left.\frac{dg_i}{d\ln\mu}\right|_{\rm SM}
 &=&
 -b_i\frac{g_i^3}{16\pi^2}~, 
 \qquad
 \{b_i\} = \left\{ -\frac{41}{10}~,~~\frac{19}{6}~,~~7\right\}~,
 \label{RGEgauge}
 \\ \vspace{-2.5cm}\nonumber\\
 \left.\frac{dy_t}{d\ln\mu}\right|_{\rm SM}
 &=&
 \frac{y_t}{16\pi^2}
 \left\{
  9y_t^2
  -
  \left(
   \frac{17}{20}g_1^2
   +
   \frac{9}{4}g_2^2 
   +
   8g_3^2
  \right)
 \right\}~, 
 \label{RGEtop}
 \\ \vspace{-2.5cm}\nonumber\\
 \left.\frac{d\lambda}{d\ln\mu}\right|_{\rm SM}
 &=&
 \frac{1}{\pi^2}
 \left\{
  \frac{\lambda^2}{4}
  -
  \frac{9}{80}\left(g_1^2 + 5g_2^2\right)\lambda 
  +
  \frac{27}{64}
  \left(
   \frac{3}{25}g_1^4
   +
   \frac{2}{5}g_1^2g_2^2
   +
   g_2^4
  \right)
 \right.
 \nonumber \\
 &&
 \qquad\qquad\qquad\qquad\qquad\qquad\qquad\qquad~~~
 \left.
  +
  \frac{3}{2}y_t^2\lambda
  -
  9y_t^4\
 \right\}~,
 \label{RGEhiggs}
\end{eqnarray}
where we neglect Yukawa couplings except the top Yukawa. From these RG
equations with the Gauge-Higgs condition $\lambda(1/L) = 0$, we can calculate
the RG flow of the Higgs quartic coupling. In Fig. \ref{fig:lambda}, the
result of the flow is depicted in the case of $\mu_{UV} = 1/L = $ 10 TeV.

From this figure, we find the Higgs mass can not exceed the present
experimental bound ($m_h \ge 114$ GeV) \cite{LEP} when the compactification
scale, $1/L = 1/(2\pi R)^{-1}$, is smaller than 10 TeV. We can resolve the
problem by increasing the compactification scale. However, it requires a
finer tuning of the quadratic coupling of the Higgs field in the original
GHU model, and it is not preferable (little hierarchy problem).

Therefore, we need other mechanisms to lift up the Higgs mass. A simple way
is to introduce additional bulk fermions to enhance the loop correction. For
instance, if we introduce $N_a$ adjoint fermions in the SU(3) model, the RG
equation for $\lambda$ is modified as 
\begin{eqnarray}
 \frac{d\lambda}{d\ln\mu}
  =
 \left.\frac{d\lambda}{d\ln\mu}\right|_{\rm SM}
 -
 \frac{3N_{A^+}}{\pi^2}
 \left\{g_4^4 + 2\left(\frac{g_4}{2}\right)^4\right\}~,
 \label{RGEhiggsAdj}
\end{eqnarray}
where $g_4$ is the effective bulk gauge coupling, which is assumed to be the
same as the SU(2) gauge coupling here. The RG flow including a couple of
adjoint fermions with a 1 TeV bulk mass is also shown in Fig.\ref{fig:lambda}.
Here, we neglect the modification of the RG equations of the gauge couplings,
because the effect is at higher order level and small in the flow of the
quartic coupling. This flow shows that the Higgs mass becomes large enough to
be able to satisfy the LEP bound.

\begin{figure}[t]
 \begin{center}
  \scalebox{0.8}{\includegraphics*{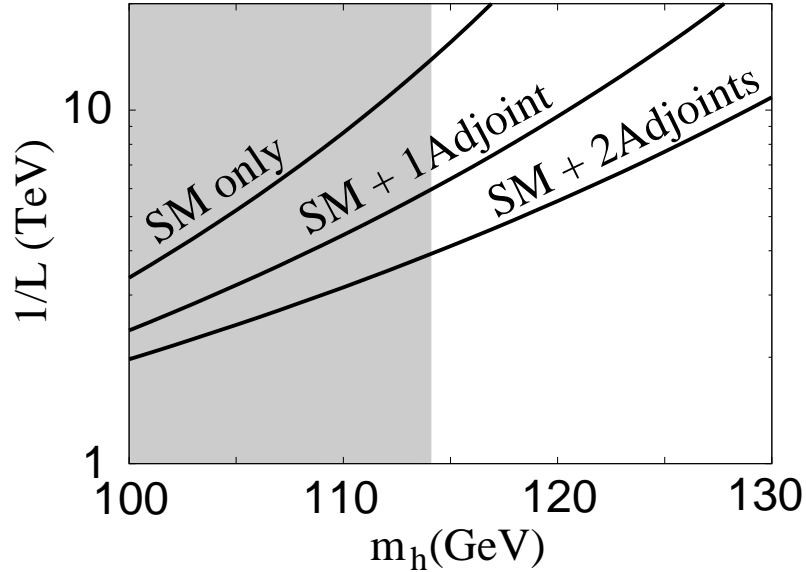}}
  \caption{\small Relation between the Higgs mass $m_h$ and the
  compactification scale $1/L$. Again the shaded area corresponds to the LEP
  bound ($m_h\ge$ 114 GeV. The upper, middle and lower lines are for
  $N_{A^+} = 0, 1, 2$, respectively.}
  \label{fig:LambdaVsHiggsMass}
 \end{center}
\end{figure}

In this way, we can evaluate the Higgs mass for a given compactification
scale, assuming the number and representation of the bulk fermions.
Conversely, we can evaluate the compactification scale from a given Higgs
mass. Results are shown in Fig.\ref{fig:LambdaVsHiggsMass}.
This kind of analysis will be useful when the Higgs mass is measured. In this
figure, it is also shown that the Higgs mass really becomes large  as the
increase of $N_{A^+}$, this fact is also suggested in the viewpoint of the
five dimensional effective potential approach \cite{HY1,higgsmass}.


\vspace{1.0cm}
\lromn 5 \hspace{0.2cm} {\bf Summary and Discussion}
\vspace{0.5cm}

We have examined a general feature of low energy effective theories in the
GHU scenario. In particular, we focus on the Higgs potential induced from the
radiative correction through the compactification. We have shown that the
low energy phenomena of GHU models can be described by the effective theory
including only the zero-modes of ingredient with the specific renormalization
condition, ``Gauge-Higgs condition''. It is surprising that the essential
informations of the GHU scenario are collected into the Gauge-Higgs condition.
It means that KK modes merely acts as regulators in the GHU scenario as far
as we examine the low energy physics.

In this letter, we have used the simple example, five dimensional SU(3) model,
for discussing the low energy effective theory. 
Even if we extend the discussion to more realistic models producing the correct
Weinberg angle and the large top Yukawa coupling and/or apply to more
complicated setup such as a six dimensional model or a GHU model in warped
extra dimension \cite{RS,warp,AdS/CFT}, 
our method would be applicable as far as the Higgs field corresponds to 
the degree of freedom of the Wilson line.
In those cases, the Gauge-Higgs condition may be 
modified from Eq.(\ref{G-H condition}). 
In fact, in the case where the shape of the extra dimension is not trivial, 
it is a non trivial question at which scale we should impose the condition.
It is, however, still expected that the running quartic coupling should 
vanish at a certain scale where the four dimensional description becomes 
inadequate. 
This is because above the scale, the Higgs field behaves as the Wilson line 
which has vanishing potential%
\footnote{
Note that when the Higgs field does not correspond to the Wilson line
\cite{6D1HDM}, there are no reasons that the quartic coupling vanishes. 
Instead, the running coupling is expected to flow toward the value of tree 
level in the original higher dimensional model.
}.

This consideration may lead to the expectation that the 2-loop corrections 
also satisfy a similar condition that includes 1-loop threshold corrections. 
This is an interesting question theoretically, but this issue is beyond the
scope of this letter and remains as a future problem. However, the higher loop
corrections are expected to be small unless we take the setup
 of strongly coupled theory nor a large number of
 baulk matter fields.

Once we clarify the general feature of effective theories, we can use them to
investigate the low energy phenomenologies of GHU models. As an example, we
have made an RG improved analysis of the Higgs mass. We have shown that some
mechanism for lifting up the Higgs mass is required in the realistic GHU
models as far as the compactification scale is less than 10 TeV. One simple
way for the lift up is to introduce bulk fermions. In this case, we
may observe some massive fermions (and no scalars !!) at future collider
experiments, even if no indications of the existence of extra-dimensions can
not be observed. Furthermore, we can observe the flow of the running quartic
coupling toward zero because of the Gauge-Higgs condition.

\vspace{1.0cm}
\hspace{0.2cm} {\bf Acknowledgments}
\vspace{0.5cm}

T.Y. would like to thank M. Tanabashi for useful discussions
 which become one of the motivations of this work.
N.H. is supported in part by Scientific Grants from 
the Ministry of Education and Science, 
Grant No.\ 16028214, No.\ 16540258 and No.\ 17740150.  
The work of S.M. was supported in paart by a
Grant-in-Aid of the Ministry of Education,
Culture, Sports, Science, and Technology,
Government of Japan, No. 16081211.
The works of N.O. are supported in part by the Grant-in-Aid for 
Scientific Research (No. 15740164) from the Ministry of Education, 
Culture, Sports, Science, and Technology of Japan.
T.Y. would like to thank the Japan Society
for the Promotion of Science for financial support.

\vspace{1.0cm}
\hspace{0.2cm} {\bf Appendix}
\vspace{0.5cm}

\vspace{0.5cm}
\underline{\bf Expansion of effective potential with bulk mass}
\vspace{0.5cm}

We show the expansion formula for the contribution to the effective potential
caused by periodic modes with bulk mass (\ref{Exact 1}). For simplicity, we
define dimensionless quantities as follow : $x = 2\pi Q_i\phi$ and $z = m/L$,
where $m$ is the bulk mass. For $x, z\ll 1$, the infinite sum in the
contribution is expanded as
\begin{eqnarray}
 &&
 \sum_{w=1}^\infty
 \frac{1}{w^5}
 \left(1 + wz + \frac{w^2z^2}{3}\right)e^{-wz}\cos(wx)
 \nonumber \\
 &=&
 \zeta_R(5) - \frac{\zeta_R(3)}{6}z^2 + \frac{1}{32}z^4
            + \frac{1}{90}z^5 - \frac{1}{1728}z^6 + {\cal O}(z^8)
 \nonumber \\
 &&
 + \frac{x^2}{2}
 \left(
  -\zeta_R(3) + \frac{7}{24}z^2 + \frac{1}{288}z^4 + {\cal O}(z^6)
 \right)
 \nonumber \\
 &&
 + \frac{x^4}{4!}\frac{1}{2}
 \left(
  \frac{25}{6} + \frac{1}{36}z^2 + {\cal O}(z^4)
 \right)
 \nonumber \\
 &&
 - \frac{1}{4!}\frac{1}{2}(z^2 + x^2)^2\ln(z^2 + x^2)
 + {\cal O}(x^6)~.
\end{eqnarray}
We can further expand the logarithm in the last line. However, the argument
$z^2 + x^2$ has the physical meaning, that is the mass of the zero-mode
normalized by $1/L$. Thus, we keep the logarithm in the above form. Then, the
infinite sum is approximated by
\begin{eqnarray}
 &&
 \sum_{w=1}^\infty
 \frac{1}{w^5}
 \left(1 + wz + \frac{w^2z^2}{3}\right)e^{-wz}\cos(wx) 
 \nonumber \\
 &\sim&
 \zeta_R(5)
 -
 \frac{x^2}{2}\zeta_R(3)
 -
 \frac{x^4}{4!}\frac{1}{2}\left\{\ln(z^2 + x^2) - \frac{25}{6}\right\}~. 
\end{eqnarray}
Comparing with Eqs.(\ref{phi4}) and (\ref{phi2}), we find that the effect of
the bulk mass modifies the argument of logarithm.

\end{document}